\documentclass{article} 
\usepackage{aaai25}  
\usepackage{times}  
\usepackage{helvet}  
\usepackage{courier}  
\usepackage[hyphens]{url}  
\usepackage{graphicx} 
\usepackage{natbib}  
\usepackage{caption} 
\usepackage{algorithm}
\usepackage{algorithmic}

\newif\ifsubmit\submitfalse

\newcommand\new[1]{\ifsubmit#1\else {#1}\fi}

\newcommand\camera[1]{\ifsubmit#1\else {#1}\fi}

\newcommand{\api}{REST API}%
\newcommand\apirl{\textsc{Apirl}}
\newcommand\rat{ARAT-RL}
\newcommand\code[1]{{\footnotesize\texttt{#1}}}
\newcommand\deeper{DeepREST}

\newcommand{\eg}{{\it e.g.}}%
\newcommand{\ie}{{\it i.e.}}%
\newcommand{\etal}{{et al.}}%
\usepackage{booktabs}
\usepackage{multicol}

\usepackage{newfloat}
\usepackage{float}
\usepackage{listings}
\floatstyle{ruled}
\newfloat{listing}{tb}{lst}{}
\floatname{listing}{Listing}

\usepackage{textcomp}

\usepackage{amssymb}
\usepackage{amsmath}
\usepackage{amsthm}
\usepackage{multirow}
\usepackage{pifont}
\usepackage[dvipsnames]{xcolor, colortbl}

\usepackage{bm}

\newcolumntype{C}[1]{>{\centering\arraybackslash}p{#1}}

\newcommand\YAMLcolonstyle{\color{red}\mdseries\footnotesize}
\newcommand\YAMLkeystyle{\color{black}\bfseries\footnotesize}
\newcommand\YAMLvaluestyle{\color{blue}\mdseries\footnotesize}

\makeatletter

\newcommand\language@yaml{yaml}

\expandafter\expandafter\expandafter\lstdefinelanguage
\expandafter{\language@yaml}
{
  keywords={true,false,null,y,n},
  keywordstyle=\color{darkgray}\bfseries\scriptsize,
  basicstyle=\YAMLkeystyle\ttfamily\fontsize{7}{7.25}\selectfont,                                 
  sensitive=false,
  comment=[l]{\#},
  morecomment=[s]{/*}{*/},
  commentstyle=\color{purple}\ttfamily\scriptsize,
  stringstyle=\YAMLvaluestyle\ttfamily\scriptsize,
  moredelim=[l][\color{orange}]{\&},
  moredelim=[l][\color{magenta}]{*},
  moredelim=**[il][\YAMLcolonstyle{:}\YAMLvaluestyle\scriptsize]{:},   
  morestring=[b]',
  morestring=[b]",
  literate =    {---}{{\ProcessThreeDashes}}3
                {>}{{\textcolor{red}\textgreater}}1     
                {|}{{\textcolor{red}\textbar}}1 
                {\ -\ }{{\mdseries\ -\ }}3,
}
\colorlet{punct}{red!60!black}
\definecolor{delim}{RGB}{20,105,176}
\colorlet{numb}{magenta!60!black}

\lstdefinelanguage{json}{
    basicstyle=\small\ttfamily\fontsize{7.2}{7.8}\selectfont,
    numbers=left,
    stepnumber=1,
    showstringspaces=false,
    breaklines=true,
    literate=
     *{0}{{{\color{numb}0}}}{1}
      {4}{{{\color{numb}4}}}{1}
      {5}{{{\color{numb}5}}}{1}
      {6}{{{\color{numb}6}}}{1}
      {7}{{{\color{numb}7}}}{1}
      {8}{{{\color{numb}8}}}{1}
      {9}{{{\color{numb}9}}}{1}
      {:}{{{\color{punct}{:}}}}{1}
      {,}{{{\color{punct}{,}}}}{1}
      {\{}{{{\color{delim}{\{}}}}{1}
      {\}}{{{\color{delim}{\}}}}}{1}
      {[}{{{\color{delim}{[}}}}{1}
      {]}{{{\color{delim}{]}}}}{1},
}

\lst@AddToHook{EveryLine}{\ifx\lst@language\language@yaml\YAMLkeystyle\fi}
\makeatother

\newcommand\ProcessThreeDashes{\llap{\color{cyan}\mdseries-{-}-}}

\usepackage{etoolbox}
\usepackage{pgf}

\renewcommand{\paragraph}[1]{{\vskip 4pt \noindent\textbf{#1.} }}

\definecolor{lowRR}{HTML}{ffffff}
\definecolor{highRR}{HTML}{0033cc}

\definecolor{lowI}{HTML}{0066cc}
\definecolor{highI}{HTML}{e6f3ff}
\definecolor{notFoundI}{HTML}{e6f3ff}

\newcommand*{\opacity}{75}

\newcommand*{\minvalRR}{6.5}
\newcommand*{\maxvalRR}{8.47}

\newcommand{\gradientRR}[1]{
    \ifdimcomp{#1pt}{>}{\maxvalRR pt}{#1}{
        \ifdimcomp{#1pt}{<}{\minvalRR pt}{#1}{
            \pgfmathparse{int(round(100*(#1/(\maxvalRR-\minvalRR))-(\minvalRR*(100/(\maxvalRR-\minvalRR))))}
            \xdef\tempa{\pgfmathresult}
            \colorbox{highRR!\tempa!lowRR!\opacity}{#1} 
            } 
        }
}

\usepackage{newfloat}
\usepackage{listings}
\DeclareCaptionStyle{ruled}{labelfont=normalfont,labelsep=colon,strut=off} 
\floatstyle{ruled}
\newfloat{listing}{tb}{lst}{}
\floatname{listing}{Listing}
\title{\apirl: Deep Reinforcement Learning for REST API Fuzzing}
\author{
    Myles Foley\textsuperscript{\rm 1}\textsuperscript{\rm 2}, Sergio Maffeis\textsuperscript{\rm 1}
}
\affiliations{

\textsuperscript{\rm 1}Department of Computing, 
    Imperial College London\\ 
\textsuperscript{\rm 2}The Alan Turing Institute\\ 
    m.foley20@imperial.ac.uk, sergio.maffeis@imperial.ac.uk
}

\begin{document}

\maketitle

%
\begin{abstract}
\api s have become key components of web services. However, they often contain logic flaws resulting in server side errors or security vulnerabilities. HTTP requests are used as test cases to find and mitigate such issues. Existing methods to modify requests, including those using deep learning, suffer from limited performance and precision, relying on undirected search or making limited usage of the contextual information. In this paper we propose \apirl, a fully automated deep reinforcement learning tool for testing REST APIs. A key novelty of our approach is the use of feedback from a transformer module pre-trained on JSON-structured data, akin to that used in API responses. This allows \apirl\ to learn the subtleties relating to test outcomes, and generalise to unseen API endpoints. We show \apirl\ can find significantly more bugs than the state-of-the-art in real world REST APIs while minimising the number of required test cases. We also study how reward functions, and other key design choices, affect learnt policies with a thorough ablation study.


\end{abstract}


\section{Introduction}


REpresentational State Transfer (REST) APIs have become the standard way to interact with Web services and resources. 
They are used by organisations such as Amazon, Google, and OpenAI to integrate with a wide array of 
systems, processes, and resources. 
These APIs consist of operations, specified by a Uniform Resource Locator (URL) accessible by the Hyper Text Transfer Protocol (HTTP), and a series of associated parameters.
Atlidakis \etal~\shortcite{atlidakis_restler_2019} show bugs in these operations are typically hard to find, 
due to the complex interactions, and diversity of APIs.
Some bugs have security implications that result in the extraction of information or data manipulation, as summarised in the OWASP API Security Top 10~\shortcite{owasp_owasp_2023}.
Hence, it is crucial to test the robustness of \api s using sophisticated techniques.


\new{
Automated software testing finds bugs or vulnerabilities in an application by detecting abnormal behaviour.
This automated process can be referred to as fuzz testing (fuzzing), security testing, or robustness testing.
%
For \api s testing, this involves creating new HTTP request test cases or mutating existing templates. 
Often, this is done at random or following some predefined heuristics.
%
%
A common criterion to practically estimate performance is code coverage.
%
However, as only a small portion of code contains bugs, it can lead to a high number of executed tests.
%
%
%

}

Black-box testing has been applied to automatically generate test cases for \api s~\cite{atlidakis_pythia_2020,liu_morest_2022}. 
While this has lead to improvements in \api\ testing, \new{such approaches lack targeted search strategies, or do not harness contextual information.}
%
This limits the potential of frameworks due to the unique behaviour of endpoints, their diversity, and varying schema requirements.
As a result software testing models \new{can be inefficient}, requiring a very large number of test cases to find bugs.

%
\new{
Recent research used attention-based neural networks to predict mutations in test cases~\cite{lyu_miner_2023}.
This is a promising approach for \api\ testing. 
Yet, current solutions often only use simplistic feedback from HTTP status codes to determine success, while the main body of HTTP responses is either discarded, or only used for populating data when testing~\cite{corradini_automated_2022,liu_morest_2022}.

%

Reinforcement Learning (RL) has shown potential in automated testing, and has been successful learning policies to test compilers \cite{li_alphaprog_2022} and web applications \cite{lee_link_2022,zheng_automatic_2021}.
These works have demonstrated that using off-the-shelf RL methods such as Deep Q-Networks (DQN)~\shortcite{mnih_human-level_2015} and Proximal Policy Optimisation (PPO)~\shortcite{schulman_proximal_2017} can harness feedback to find more bugs, and improve efficiency.
However, similar to \api\ testing, RL-based software testing often uses simple heuristics as feedback~\cite{lee_link_2022,li_alphaprog_2022}, which may fail to fully capture the subtleties of the problem.
%
While work from Kim \etal~\shortcite{kim_adaptive_2023} has shown that RL can find bugs in \api s, it suffers from the same issue: not harnessing contextual information for feedback.
The key challenge to do so arises from the variation across different APIs and the diversity in individual endpoint responses.}
Yet, if the information-rich data structure can be used for test case generation, it could provide valuable feedback.
To address this challenge, we develop a novel deep RL agent that mutates HTTP requests to find bugs in \api s. 

Our RL problem formulation uses a transformer architecture we pre-train to harness JSON and natural language in the HTTP responses, providing feedback to facilitate adaptation to operations after training.
We then introduce a Markov Decision Process (MDP) for the mutation of HTTP requests as a number of sequential changes from an initial HTTP request (referred to as \emph{request template}).
Between mutations, the new test cases are submitted to an API, using the response code and execution information to determine the reward used in training. 
\new{The trained policy uses the mutation strategy to find bugs in unseen APIs with a minimal number of HTTP requests, overcoming a significant limitation of web based testing approaches~\cite{lee_link_2022}.
%
}
%
%

In summary, the main contributions of this paper are: 
\begin{itemize}
    \item \apirl, a novel \new{deep} reinforcement learning based approach to testing \api s that learns how to manipulate varied HTTP requests to efficiently target bugs. We release \apirl\ at \url{https://github.com/ICL-ml4csec/APIRL}.
    \item A novel representation of the feedback obtained from the API, combining a direct functional representation of fixed outputs and a transformer-based embedding of variable-length responses. This enables our RL agent to benefit from richer information than in previous work.

    \item \new{
    Insights on the subtleties of training an RL agent for real-world tasks via an ablation study of both design choices and 7 reward functions.
    } 
    
    \item The evaluation of \apirl\ \new{across 26 \api s} shows significant improvement over state-of-the-art methods in terms of bugs found, coverage, and test case efficiency.
    

\end{itemize}

\section{Background}\label{sec:background}

\subsection{Deep Q-Networks}\label{sec:rl_background}


%
%
%
%

\new{
A \emph{Deep Q-Network} (DQN) leverages deep neural networks to learn an \emph{optimal} policy ($\pi^*$). 
In the DQN a Q-network, with parameters \bm{$\theta$}, computes Q-values $q_{\pi}(s, \cdot, $\bm{$\theta$}$)$ 
corresponding to actions $a \in A$ given a state  $s \in S$.
Actions are taken via $\varepsilon$-greedy selection, where random actions occur with probability $\varepsilon$, otherwise taking the `greedy' action dictated by the policy.
The state then transitions to a new state $s^\prime \in S$.
%
%
A Target Q-network, with parameters \bm{$\theta^-$} predicts the Q-value of this next state $q_{\pi}(s_{t+1}, \cdot, $\bm{$\theta$}$)$.
Then using semi-supervised approach the loss is computed from the Q-value predictions of the  Q-network and Target Q-network using the Bellmen equation~\cite{sutton_reinforcement_2018}.
%
%
During training, the parameters \bm{$\theta^-$} are periodically updated with \bm{$\theta$}. 
This allows for the convergence of the two networks to the optimal policy \(\pi^*\)~\cite{silver_mastering_2016}. 
\emph{Prioritised experience replay}~\cite{schaul_prioritized_2016} improves training efficiency by weighting samples by their loss, increasing their sample probability. Importance sampling then assigns weights to transitions to remove the bias from the change in sampling.

}

\subsection{OpenAPI Specification}

The OpenAPI Specification~\shortcite{noauthor_home_nodate} is the standard for describing \api s. It specifies the URL and HTTP method (\code{POST}, \code{GET}, \code{PUT}, \code{PATCH}, and \code{DELETE}) to form an \emph{operation}.
Each operation includes the schema for associated parameters, 
and any required authentication. 
%
Figure~\ref{fig:vapi_openapi} shows a snippet of the specification for 
an VAmPI~\shortcite{erev0s_vampi_2023}. 
This example includes the \code{username} parameter in the URL and key-value pair in the request body of \code{\{"email": "value"\}}.
Each parameter is assigned a name, type, and can contain details such as examples, or even if it is required in the operation. 
%
The HTTP response code specified in Figure~\ref{fig:vapi_openapi} of `204', indicates a successful request. However the response could be in the range 400-499 (4XX) indicating a user error, or 500-599 (5XX) indicating a server error or bug.

\section{Overview}

\begin{figure*}
    \centering
    \includegraphics[width=0.9\textwidth]{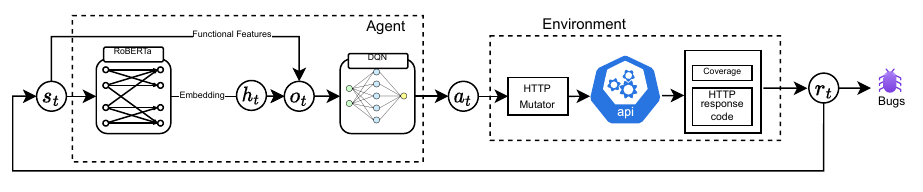}
    \caption{The \api\ testing process using \apirl.}
    \label{fig:apirl_design}
\end{figure*}

\apirl\ is a new testing approach based on deep Q-learning that mutates HTTP requests to find bugs in \api s, indicated by 5XX response codes.  
We represent \api\ testing as an MDP for a deep RL agent:  Figure~\ref{fig:apirl_design} shows the process and agent architecture.  
At a high level, an agent takes actions to mutate HTTP \emph{request templates}, which the environment implements as API operations, receiving feedback that forms the reward.
\apirl\ takes a HTTP request-response pair as the input state $s_t$.
Maximal information is then extracted from functional features and an embedded representation \new{via a pre-trained transformer (Section~\ref{ssec:obs}).
Using diverse and variable length features such as HTTP headers and body data, a neural network selects a corresponding action $a_t$ from Table~\ref{table:actions} to mutate the HTTP request.}
%

Using the agent-selected action, the environment then performs concrete mutations on the HTTP request template forming \emph{test cases}. 
The model evaluates the test case performance using the HTTP status code and the execution trace of the \api. 
\new{We develop these to form varied reward functions to study their effect in training (Section~\ref{ssec:rewards} and~\ref{ssec:ablations}).}
%
%
The model evolves to optimise for the reward as it performs more mutations, and performance evaluations. 
%


We compare the learnt policy against a state-of-the-art \new{learning and non-learning tools. Finally, in an ablation study we evaluate seven reward functions and key design choices.} 
%
%

\begin{figure}[t!]
\noindent\begin{minipage}{\textwidth}
\begin{lstlisting}[language=yaml,frame=single,linewidth=0.45\textwidth,numbers=none]
"/users/v1/{username}/email":
    put:
      parameters:
        - name: username
          in: path
          required: true
          type: string
      requestBody:
        content:
          application/json:
            schema:
              type: object
              properties:
                email:
                  type: string
        required: true
      responses:
        '204':
          content: {}
\end{lstlisting}
\end{minipage}
\caption{Part of the OpenAPI specification for VAmPI}\label{fig:vapi_openapi}
\end{figure}

\section{Design}\label{sec:approach}

In this section, we define \new{our} RL-based \api\ fuzzing \new{approach} and detail the model design of \apirl.
%

\subsection{Preprocessing}
%

The OpenAPI specification is the standard starting point for \api\ fuzzing frameworks~\cite{liu_morest_2022,atlidakis_restler_2019}. 
From a specification, we extract the operations and their associated parameters. This forms a list of request templates (HTTP method, URL, parameters details) \new{for the RL agent} to mutate. For example, the operation in Figure~\ref{fig:vapi_openapi} would have a request template such as: 
(\code{PUT}, \code{/users/v1/\{username\}/email}, \code{\{name:username, in:path, required:true, type:string\}}).
%
We also find related parameters to form valid requests, as recommended by the literature~\cite{martin-lopez_specification_2021}. 
For more on this see Appendix~\ref{app:key-value-pairs}. 

\subsection{Actions}

We provide the RL agent with a fixed action space of 23 actions $a_t$ used in prior work~\cite{atlidakis_restler_2019,barabanov_automatic_2022,corradini_automated_2022}.
%
\camera{We reimplement actions in the \apirl\ framework using distinct values where possible, otherwise we provide concrete definitions (Table~\ref{table:actions}).}
%
%
Actions are performed on requests, independently of specific applications so \apirl\ does not need further training or feedback from rewards when encountering new \api s, thus aiding generalisation.


Actions are detailed in Table~\ref{table:actions}, and can be broadly devided in three categories. 
Actions 1 and 2 that alter the authorisation token by either refreshing the current authorization token (if an endpoint allows for this), or switching to an alternative authorization token if one has been provided. 
%
Action 3, that allows the agent to switch to begin mutating the next parameter for this operation. This then loops to the start of the parameters upon reaching the end. 
Finally, actions 4-23 that manipulate the request template to perform \api\ fuzzing functionality. We utilise those that may trigger bugs by duplicating or removing parameters, using default values, and finding alternative endpoints.

Using actions in Table~\ref{table:actions} the agent mutates a request at each timestep. After a mutation the request is sent to the \api\ and\new{, in training,} runtime information is collected for the reward (Section~\ref{ssec:rewards}). \new{The episode then terminates if a bug is found (as indicated by a 5XX status code) or if the maximum number of timesteps have elapsed.}



\subsection{Observations}\label{ssec:obs}
%

%

Making use of information from real world tasks for RL observations can be challenging due to the complexity, high dimensionality, or format (such as natural language).
%
We aim to make \apirl\ bridge this semantic gap with a pre-trained RoBERTa~\cite{liu_roberta_2019} transformer model. 
%
%
\apirl's transformer takes as input the HTTP response from test cases to produce a latent representation. \apirl\ can then harness complex JSON and natural language from the structure, text, headers, and encoding of API responses.
\camera{The latent representation forms a vector of 768 features from the RoBERTa transformer (its standard output feature dimension), $h_t$ in Figure~\ref{fig:apirl_design}.
The observation further includes: 
the HTTP method of the current operation, 
the HTTP response code, 
the variable type of the parameter currently being manipulated, 
and the normalised index of this parameter out of all parameters in the request template.
These features are represented as a vector of length four.
Both vectors are combined ($o_t$ in Figure~\ref{fig:apirl_design}) into a single vector of 772 (768 + 4) features and passed to the DQN. 
See Section~\ref{sssec:training} for further details on pre-training the transformer.}

\begin{table*}[t!]\centering
\caption{Mutation actions that can be applied by \apirl\ to the current operation.}\label{table:actions}
\resizebox{\textwidth}{!}{
\begin{tabular}{@{}lcll@{}}
\toprule
Action Type & \begin{tabular}[c]{@{}l@{}}Action \\ Number\end{tabular} & Example & Description \\ \midrule
Auth Token Refresh & 1 &  & Refresh the any authorization token. \\
Switch Auth Token & 2 &  & Use the authorization of another user. \\
Switch Parameter & 3 &  & Change to the next parameter in request template. \\
Change Parameter Type & 4 & \begin{tabular}[c]{@{}l@{}}\texttt{Int}, \texttt{String}, \texttt{Bool}, \texttt{Array}, \\\texttt{Object}\end{tabular} & \begin{tabular}[c]{@{}l@{}}Randomly change the parameter value type to a \\different one.\end{tabular} \\
Duplicate Parameter & 5 & \texttt{\{parameter1:[value1, value1]\}} & Duplicate the parameter value. \\
Remove Parameter & 6 & \texttt{\{\}} & Remove the parameter from the request template. \\
Extension & 7- 9 & \texttt{.txt}, \texttt{.pdf}, \texttt{.doc} & Append a file extension to the value of the parameter. \\
Append & 10 & \texttt{\{parameter1:[value1, newValue]\}} & \begin{tabular}[c]{@{}l@{}}Convert the parameter to a JSON array and append \\ an additional value from observed values.\end{tabular} \\
Request Method & 11 & \texttt{POST} $\rightarrow$ \texttt{PUT}, \texttt{PUT}$\rightarrow$ \texttt{POST} & Change HTTP method from \texttt{PUT} to \texttt{POST} or \texttt{POST} to \texttt{PUT}. \\
Add Parameter & 12 & \begin{tabular}[c]{@{}l@{}}\texttt{\{parameter1:value1,}\\  \texttt{ newParameter:newValue\}}\end{tabular} & \begin{tabular}[c]{@{}l@{}}Add an additional parameter to the request template \\ from parameters related to the parameter.
\end{tabular} \\
Wildcard & 13-15 & \texttt{*}, \texttt{.*}, \texttt{\%} & Append a wildcard to the value of the parameter. \\
Change ID number & 16-17 & \texttt{1}, \texttt{-1} & Increment the \texttt{Int} value of the value by 1. \\
Set Parameter value & 18-21 & \texttt{'admin'}, \texttt{-1}, \texttt{999999999} ,\texttt{''} & Set the parameter value to a default value. \\
Set Existing Value & 22 & \texttt{\{parameter1:value2\}} & Set the parameter value to a related value from the API. \\
Set admin & 23 & \texttt{\{parameter1:value1, admin:TRUE\}} & Set \texttt{admin: True} in an \texttt{Object}. \\ \bottomrule
\end{tabular}}
\end{table*}

\subsection{Reward}\label{ssec:rewards}

Two main signals can be used to guide \api\ testing: 
a) coverage of the \api; and b) the HTTP response code of the request. 
Coverage reflects the ability to explore the API behaviour, blindly aiming to exercise as much of the implementation code as possible, hoping to trigger bugs in the process. On the other hand, the HTTP response code of the request provides information to guide fuzzing, including the validity of requests in terms of authorisation (\eg\ 401), parameters (\eg\ 200 or 400), and server-side errors or bugs (\eg\ 500). 
\new{We define the reward for \apirl\ based on the HTTP response code as it provides more nuanced feedback on test case performance (Eq.~\ref{eq:r_sc}). 
We will consider alternative rewards, including coverage, in an ablation study (Section~\ref{ssec:ablations}).

\begin{equation}\label{eq:r_sc}
    R_{sc} =
    \begin{cases}
     10, & \parbox[t]{.25\textwidth}{$5XX$ HTTP status response} \\
     1, & \parbox[t]{.25\textwidth}{$2XX$ HTTP status response} \\
     -1, & \parbox[t]{.19\textwidth}{Otherwise} \\
    \end{cases}
\end{equation}

$R_{sc}$ incentivises the agent most for the desired behaviour of finding bugs on the server-side of the \api. However, as this can be a sparse reward, we provide interim feedback 
for performing correct requests.
In all other cases we discourage the behaviour using a negative reward~\cite{sutton_reinforcement_2018}.
This reward is \emph{consistent} as training progresses so the agent learns to develop diverse requests, covering more of the back-end of the \api.

}

\subsection{Agent Architecture}\label{ssec:agent}

To develop mutational strategies that can be dynamically altered to specific operations and \api s, we develop a deep RL agent based on the Deep Q-Network (DQN)~\cite{mnih_human-level_2015} with \emph{Prioritised experience replay}~\cite{schaul_prioritized_2016}. We implement the neural network in PyTorch, with an input layer of size 772, and  hidden layers of size 64, 96, 64, with an output layer of 23, corresponding to the actions in Table~\ref{table:actions}. We use the ReLU activation function after each hidden layer. 
The agent learns which mutation, or combination of mutations to apply, while reducing the computational complexity of the neural network. We use a standard  $\varepsilon$-greedy decay with $\varepsilon=1$ (decaying by $0.999$ after each episode). We select $\gamma = 0.9$, $\alpha = 0.005$, and batch size 128. These are selected via gridsearch, further details and parameter values can be found in 
Appendix~\ref{app:hyperparam}.

\section{Evaluation}\label{sec:experiments}

\subsection{Training}\label{sssec:training}

Training should result in an RL agent that can manipulate HTTP requests, generalising to find bugs in different APIs. 
\new{Thus, we train} \apirl\ on multiple endpoints, targeting each operation 
for a set number of episodes. 
\new{This form of curriculum learning prevents over-fitting by dividing training equally across diverse operations~\cite{wahaibi_sqirl_2023}.}
Specifically, \apirl\ is trained using an open-source \api\ containing known bugs: Generic University~\cite{paxton-fear_generic_2023}.
\apirl\ trains on each of Generic University's operations for 10,000 episodes, \new{with the maximum steps per episode of $10$ (Appendix~\ref{app:hyperparam})}.

We pre-train a RoBERTa transformer using HTTP responses from 103 different \api s, comprising 1283 operations. This has several advantages: limiting the bias from HTTP responses in training, potential for overfitting, and reduces instability in training the RL agent~\cite{parisotto_stabilizing_nodate}.
API specifications were taken from a public OpenAPI specification platform (see Appendix~\ref{app:data}).
A vocabulary of 52,000 tokens is formed via Byte-Pair Encoding~\cite{gage_new_1994}. The transformer is trained by masked language modeling~\shortcite{devlin_bert_2019}.
\new{Through training the transformer learns relationships between parameters, to provide a meaningful, generalised, embedding for the agent~\cite{adolphs_ledeepchef_2019}.}
A gridsearch is used to select hyperparameters as in Appendix~\ref{app:hyperparam}.
%

\new{

As the state-action space is target agnostic, the trained policy can be used on unseen \api s without the high number of training iterations required to reach optimal performance.
An advantage of this behaviour is no further learning or feedback for the reward (\eg\ code coverage) is required. As such we run \apirl\ in black-box fashion. 
}

\begin{table*}[!t]\caption{
Average unique bugs, LoC, and number of requests taken on each test service, with standard deviation shown in brackets. 
}\label{tab:results}\centering\resizebox{\textwidth}{!}{
\begin{tabular}{@{}cccccccccc@{}}
\toprule
\rotatebox[origin=c]{55}{Test Service} & \rotatebox[origin=c]{55}{Operations} & \rotatebox[origin=c]{55}{Metric} & \rotatebox[origin=c]{55}{EvoMaster} & \rotatebox[origin=c]{55}{RestTestGen} & \rotatebox[origin=c]{55}{\deeper} & \rotatebox[origin=c]{55}{\rat} & \rotatebox[origin=c]{55}{MINER} & \rotatebox[origin=c]{55}{Rand-\apirl} & \rotatebox[origin=c]{55}{\apirl} \\ \midrule
\multirow{3}{*}{Capital} & \multirow{3}{*}{16} & Unique Bugs & 0 (0) & 0 (0) & 0 (0) & 0 (0) & 0 (0) & 0 (0) & 0 (0) \\
 &  & LoC & 500.2 (21.7) & 592.6 (9.9) & 596.8 (19.6) & 385.8 (74.2) & 446 (0) & 727.4 (11.3) & \textbf{751.3 (11.1)} \\
 &  & Requests & 1670.2 (310.7) & 1074.6 (0.5) & 1075.8 (1.1) & 3437.8 (3574.9) & 15785.4 (1113.5) & \textbf{680 (0)} & \textbf{680 (0)} \\ \midrule
\multirow{3}{*}{VAmPI} & \multirow{3}{*}{13} & Unique Bugs & 1.4 (1.1) & 2.4 (0.5) & 2.4 (1.5) & 2 (1.4) & 2 (0) & 1.8 (1.3) & \textbf{3 (0.7)} \\
 &  & LoC & 245.8 (83.2) & 327.6 (42.8) & 278 (9.1) & 272.4 (10.4) & 277 (0) & 327.2 (51.4) & \textbf{382.8 (8.7)} \\
 &  & Requests & 4564.6 (1190.6) & 3990.6 (874.6) & 5056.6 (1651.3) & 2723 (2072.8) & 10701.2 (150.7) & 526.2 (11.8) & \textbf{521 (4.4)} \\ \midrule
\multirow{3}{*}{vAPI} & \multirow{3}{*}{17} & Unique Bugs & 5.5 (0.6) & 6.4 (0.5) & 7.2 (0.8) & 2.2 (1.5) & 6 (0) & 7.6 (1.1) & \textbf{9.5 (0.6)} \\
 &  & LoC & 362.2 (1.6) & 349 (0) & 359.8 (6.6) & 359.2 (17.9) & 344 (0) & 362.8 (16.5) & \textbf{385.7 (6.5)} \\
 &  & Requests & 629.75 (41.3) & 1330.6 (68.5) & 520.8 (249.6) & 1375.6 (53.1) & 1952.8 (500.7) & 394.6 (70.4) & \textbf{243.75 (14.2)} \\ \midrule
\multirow{2}{*}{Spree} & \multirow{2}{*}{68} & Unique Bugs & 1.7 (1.5) & 0 (0.0) & 0.5 (0.6) & 0.3 (0.6) & 1.75 (1.5) & 2.25 (1.0) & \textbf{10.2 (0.8)} \\
 &  & Requests & 1657.0 (99.0) & 2034.2 (1.0) & 2439 (396.0) & 2169.3 (232.0) & 16019.75 (2802.1) & 1459.5 (78.0) & \textbf{1128.6 (37.6)} \\ \midrule
\multirow{2}{*}{BitBucket} & \multirow{2}{*}{518} & Unique Bugs & 6.2 (1.3) & 0 (0) & 3.6 (4.2) & 3.4 (4.3) & 6.3 (1.5) & 5 (1) & \textbf{6.6 (1.1)} \\
 &  & Requests & 1315.8 (438.6) & 1194 (0) & 3168.2 (276.8) & \textbf{1079.4 (595.3)} & 1485.25 (46.5) & 10593.6 (11.9) & 10562.8 (7.5) \\ \midrule
\multirow{3}{*}{WordPress} & \multirow{3}{*}{191} & Unique Bugs & 0.6 (1.3) & 1.2 (0.4) & 0 (0) & 1 (1.0) & 0.6 (0.5) & 3 (2.8) & \textbf{5.2 (3.5)} \\
 &  & LoC & 5658.5 (514.1) & 7843.8 (105.8) & 7075.2 (29.7) & 5295 (0.0) & 5639.8 (331.6) & 7995 (159.5) & \textbf{8122.2 (117.5)} \\
 &  & Requests & 6327.6 (5138.8) & \textbf{2298.4 (35.3)} & 5835.4 (125.3) & 5064.8 (889.7) & 7373.6 (3380.3) & 5685.3 (25.9) & 5678.4 (55.4) \\ \bottomrule
\end{tabular}}
\end{table*}

\subsection{Experiment Setup}
To test \apirl\ we make several modifications to its setup: we limit the number of episodes per operation to three, and reduce the probability of taking random actions ($\varepsilon$) to 5\%. 
This was seen to balance runtime and bug finding by reducing wasted requests on true negatives or invalid actions. 
Experiments are run on Ubuntu Linux, with 16GB RAM and Intel core i7 8700k processor. To mitigate the intrinsic stochasticity of approaches, we repeat each experiment five times. 

\paragraph{\api s} 
We evaluate our framework on smaller \api s, including: VAmPI, vAPI v1.3, and c\{api\}tal. 
%
\new{We conduct large scale tests for bugs in APIs from large-scale projects. Spree Commerce v4.4.0 (a popular e-commerce platform with 12.5k stars on github) containing 2 APIs, 17 APIs from BitBucket v8.2.1 (a popular git hosting service with over 15 million users),  and 4 APIs from WordPress v6.6.1 (an opensource web platform used in over 5 million websites).} These \new{26 APIs} represent \new{823} separate operations. 
We run tests on VAmPI, vAPI, and c\{api\}tal for a maximum of 1.5 hours, and 10 hours on BitBucket,  Spree Commerce \new{and Wordpress} APIs, due to their complexity and increased number of operations.
\camera{Testing with such time limits is in line with other studies~\cite{liu_morest_2022,kim_automated_2022}. Additionally, broader studies in fuzzing, such as by B\"ohme \etal~\shortcite{bohme_coverage-based_2016} have shown that most fuzzers find the majority of coverage early in testing, after which they asymptomatically converge.}
Due to ethical concerns over data privacy, integrity, and availability, \new{\emph{we do not test on live \api s}}, but instead deploy them \new{locally}. 
Each \api\ is initialised with \new{\emph{non-sensitive}} data. After each test we restore it to the state prior to testing to remove any bias.

\paragraph{Baselines} 
In order to compare with the state of the art, we select six black-box baselines.  
%
\textit{MINER}~\shortcite{lyu_miner_2023} builds call sequences using feedback from executed requests and an attention based neural network to generate mutational parameters. 
\new{\textit{\rat}~\shortcite{kim_adaptive_2023} is a tabular-Q-learning based \api\ tester.
\textit{\deeper}~\cite{corradini_deeprest_2024} is a  mulit-learning approach using an LLM to generate key value pairs, a Deep RL agent to select operations, and a series of Multi-Armed Bandits (MABs) per parameter to select mutations.
\textit{EvoMaster}~\shortcite{arcuri_automated_2021} is based on evolutionary algorithms and heuristics.
\textit{RestTestGen}~\shortcite{corradini_resttestgen_2022} uses the specification to generate Operation Dependency Graphs (ODGs) to create call sequences. } 
%
We also use \textit{Rand-\apirl}, a variant of our approach that selects actions at random. 
%
\camera{We also provide each tool with the required authentication token. 
}

\paragraph{Evaluation Criteria}
We use several different criteria in our evaluation.
\textit{Line coverage} (LoC) measures precise coverage performance. Details on how this is collected can be seen in Appendix~\ref{app:code}.
Note that coverage cannot be collected from 
BitBucket as it is closed-source \cite{liu_morest_2022}, and Spree Commerce does not support coverage collection across its multi-server architecture.
We use the \textit{request volume} \new{as recommended by Caturano \etal~\shortcite{caturano_discovering_2021}. It represents how `intelligent' an approach is, where better scanners achieve similar results in fewer requests. }
%
The \textit{number of bugs} found in \api s is a key criteria and the goal of testing.
Achieving high coverage of \api s may trigger logic flaws or server-side errors, as indicated by a 
5XX status code.
We define unique bugs as 5XX responses from each operation that originate from different lines of source code. 
\new{
Detecting bugs this way provides a simple and effective oracle within the scope of this work~\cite{arcuri_automated_2021}.

\new{The experiment results are shown in Table~\ref{tab:results}, and the average improvement of \apirl\ over the respective baselines is shown in Figure~\ref{fig:results}.}
}

%


\subsection{Coverage and Request Volume}\label{ssec:RQ1}

Coverage is commonly used to determine tester performance, with higher coverage increasing test completeness to find more bugs.
However, we also consider the \emph{efficiency} of approaches \new{in terms of the requests used}.
%
%

%

\begin{figure*}
    \centering
    \includegraphics[width=0.7\textwidth]{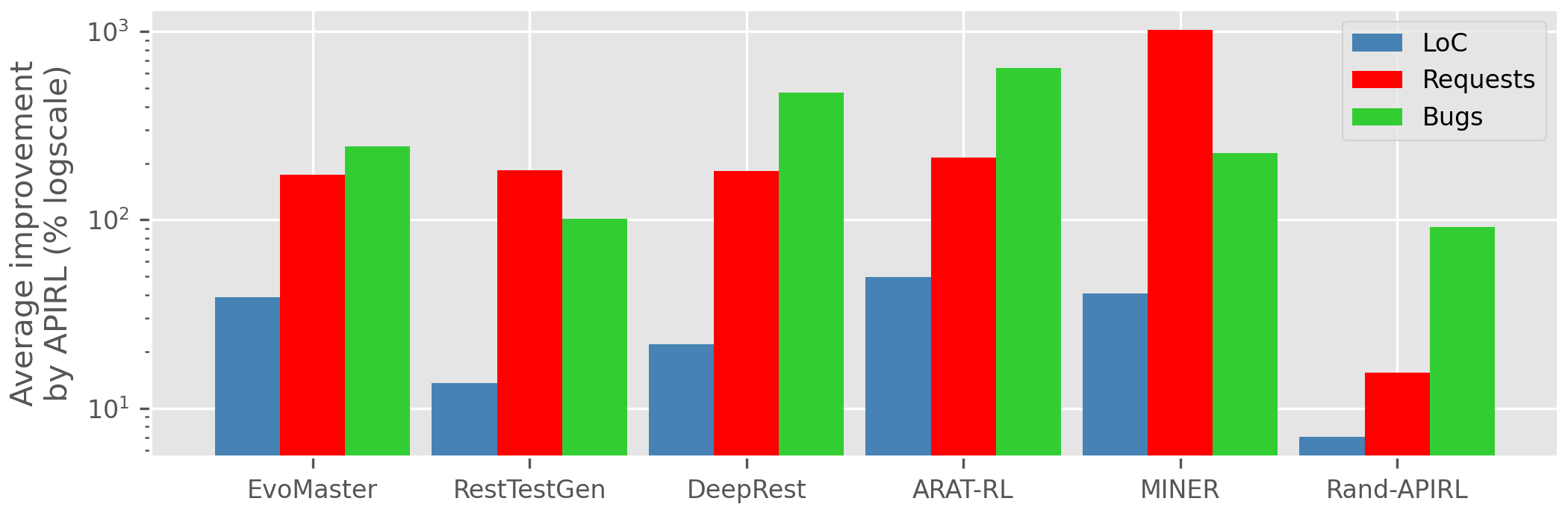}
    \caption{Improvements of \apirl\ compared to baselines.}
    \label{fig:results}
\end{figure*}
\new{\apirl\ outperforms the random baseline in all cases, covering at least 7.0\% more code, using 15.4\% fewer requests. The performance distributions differ between \apirl\ and Rand-\apirl, with no overlap in their first standard deviation.}
\apirl\ always uses the same number of requests in c\{api\}tal as no bugs were found, resulting in all episodes reaching maximum length.
\new{MINER consistently achieves low coverage; where a lack of targeted mutations is apparent from the significantly high number of requests in both Table~\ref{tab:results} and Figure~\ref{fig:results}.
While \rat\ and RestTestGen uses minimal requests in two case studies their performance is inconsistent, as is their coverage performance.
The evolutionary approach of EvoMaster reduces requests compared to heuristic counterparts. Yet, on average, it requires 173.8\% more requests than \apirl, resulting in 38.9\% less coverage. 
\deeper s multi-learning approach doesn't impact the rate of requests it can send as it achieves a high number of requests when testing endpoints. 
Yet, \deeper\ is unable to use this to find increased coverage, having a middling performance achieving 13.9\% less coverage than \apirl, using 81.9\% more requests.
Overall, \apirl\ improves over RestTestGen, \rat, and MINER by 13.6\%, 49.7\%,  40.6\% respectively in terms of coverage.
The consistent performance of \apirl\ results in higher efficiency of coverage per request, on average 64.8\% higher. 
The lower standard deviation of \apirl\ further demonstrating the \emph{consistency} in targeting mutations to improve over both Rand-\apirl\ and state-of-the-art.

}

\subsection{Bugs Found and Request Volume}\label{ssec:rq2}
We now investigate the ability to find bugs in the systems under test.
\new{\apirl\ outperforms all baselines, finding at least 6.4\% more bugs, and significantly more bugs on average (Figure~\ref{fig:results}).
\apirl\ finds these bugs with lower or competitive standard deviation. 
It has a higher standard deviation than its random counterpart only twice. 
Indeed Rand-\apirl\ has an expectedly high standard deviation, which leads to an overlap in number of bugs found in BitBucket, VAmPI, and WordPress. 
Yet, Rand-\apirl s best case performance does not outperform \apirl, which finds more bugs in fewer requests.
\rat\ displays the worst bug finding performance of all approaches, which combined with the high number of requests results in low bugs found per request.
While EvoMaster and MINER find bugs across diverse \api s, they find 117.6\% and 99.2\% fewer bugs than \apirl\ respectively. 
RestTestGen inconsistently finds bugs due to its heuristic matching of parameter names for creating the call graph ODG.
\deeper, being based on RestTestGen, suffers from similar performance issues, finding 63.8\% fewer bugs than \apirl. 
}


In BitBucket 
\camera{we see an anomaly, \apirl\ uses more requests than other baselines. 
This is due to the larger number of \api\ endpoints that do not contain bugs. 
\apirl\ is configured to have a maximum of 3 episodes and ten requests sent per episode, meaning that the upper limit of requests for the 518 operations is $3\times10\times518=15,540$.
%
%
%
In Spree we see a different story, as \apirl\ achieves the lowest number of requests of all approaches. 
}

Using our state-action space improves results compared to \new{state-of-the-art} as Rand-\apirl\ occasionally finds more bugs. 
\camera{We reason the \apirl\ test framework gives it an `edge' over the other testing strategies.}
However, \apirl\ uses the state-action space effectively, always finding more bugs in fewer test cases. Highlighting the ability of RL to efficiently target mutations, even when testing on unseen \api s.

\camera{

Of the 49 unique bugs 22.4\% were unique to \apirl\ and no other scanner. EvoMaster alone found a unique bug. 
\apirl\ misses 7 bugs, 31.8\%, as Rand-\apirl, and 68\% less than EvoMaster.
%
%
MINER, \deeper, and \rat\ find 5 of the 7 bugs missed by \apirl, yet this has an increased cost in the number of requests required, and a lower overall number of bugs found.
Such bugs can be found by including additional keys-values pairs in requests with default (and intentionally incorrect) values. 
\apirl\ can only include parameter keys from the schema so it cannot trigger this functionality.

\apirl\ can trigger handling errors by inserting \code{Int} and \code{String} objects, and removing required parameters.
%
\apirl\ learns to cast objects to alternative types, causing bugs in 7 endpoints in Spree which are missed by other approaches. 
%
%
Furthermore, \apirl\ misses no bugs in WordPress, Spree, or vAPI, . 
%
}
%
%
%
\camera{In WordPress's POST \code{wp/v2/posts/{id}}\footnote{\url{https://core.trac.wordpress.org/ticket/61837}} operation \apirl\ triggers a \code{Type} error by inserting an additional \code{password} parameter into the body of the request. Due to the incorrect type being an \code{Int} and not a \code{string}, WordPress correctly throws an error, however this results in an authentication check using the password parameter. Finally, this attempts to check the hashed POST password against the real password, which throws a fatal error.}
%
A similar bug is found in BitBucket
by duplicating a parameter value in a request such as \code{"all": [true, true]} this triggers a \code{Type} error when the \code{Array} is cast as a \code{Bool}. 
%

\new{\apirl\ has also learnt to trigger unhandled SQL errors.}
In VAmPI it inserts existing parameter values into the request that fail unique constraints. 
In Spree Commerce \apirl\ inserts an additional parameter that is included in the SQL query, leading to the query trying to access a column that does not exist.
\new{An example of the long term strategy of \apirl\ is shown by manipulating a series of \code{PATCH} requests in Spree Commerce. By bypassing input sanitisation \apirl\ causes an error by trying to access the attributes of the user. Interested readers may refer to Appendix~\ref{app:example_bug}}.

\subsection{Ablation Study}\label{ssec:ablations}
\new{
\apirl\ displays impressive performance, finding bugs in unseen \api s, in a minimal number of requests.
However, we wish to determine the extent to which core elements of \apirl\ contribute to performance.
Thus, we conduct an ablation study, presenting the results in Table~\ref{tab:ablations}.}
%
In particular, we vary rewards based on coverage (\apirl-cov) and response code (\apirl), we ablate transformer embeddings (\apirl-m), and change the RL training algorithm to PPO (\apirl-p)\new{, see Appendix~\ref{app:hyperparam} for hyperparameters}. 
\new{Interested readers may see Appendix~\ref{app:importance} for a feature importance study further confirming the utility of transformer embeddings.}

\paragraph{Reward Variations}
Finding the correct reward function is fundamentally important as it provides the feedback for learning. 
To prevent agents from finding exploitative strategies in complex reward functions we design simple reward functions~\cite{sutton_reinforcement_2018}. 
Prior work from Li \etal~\shortcite{li_alphaprog_2022} studied how reward functions alter behaviour of RL agents when generating code for compiler testing. 
%
\new{Further work from Bates \etal~\shortcite{bates_reward_2023} showed the challenges of designing rewards for RL applications in cyber-security.

Thus, we investigate reward functions for \emph{manipulation} of HTTP request templates in automated testing of \api s.
By using diverse sets of rewards based on different testing signals, we can study how subtleties in rewards influence learning policies that test real-world systems. }
While both coverage and status code provide feedback for RL agents to test \api s, it is unclear how they will affect the learning. 
As such we develop several reward variants:



\begin{itemize}
    \item \textit{A coverage based reward ($R_{cov}$)}. Similar to $R_{sc}$, $R_{cov}$ rewards most for increasing coverage (10), giving a smaller reward for recovering the same code (1), and penalises otherwise ($-1$). We refer to this agent as \apirl-cov.

    \item \textit{A sparse uniform reward}. We train \apirl-u by rewarding $1$ for 2XX and 5XX response codes and $-1$ otherwise. Equally for \apirl-cov-u we train a coverage variant that rewards $1$ for new unique LoC and $-1$ otherwise.
    \item \textit{A reward ratio}. We train \apirl-r with status code ratio of: $r=\sum(5XX+2XX)/\sum(XXX)$. We also train \apirl-cov-r using $r=\sum(LoC_{new})/\sum(LoC)$.

    \item \textit{An \rat\ style reward}.
    Similar to $R_{sc}$ Kim \etal~\shortcite{kim_adaptive_2023} design a status code based reward that gives $1$ for $4XX$ and $5XX$ responses, penalising $-1$ for $2XX$. %
    The model trained with this reward is \apirl-arat.

    %

\end{itemize}

\begin{table}[t!]\caption{Ablations of both \apirl\ and \apirl-cov. 
 Lighter colours represent poor performance, while {\color{highRR} darker colours} represent good performance.
}
\label{tab:ablations}\centering\resizebox{\columnwidth}{!}{

\begin{tabular}{@{}cccccc@{}}
\toprule
\begin{tabular}[c]{@{}c@{}}  \apirl \\ Variant \end{tabular}& \begin{tabular}[c]{@{}c@{}} Successful \\ Requests\end{tabular} & \begin{tabular}[c]{@{}c@{}} Error\\Requests\end{tabular} & \begin{tabular}[c]{@{}c@{}}Invalid\\Requests\end{tabular} & Coverage & \begin{tabular}[c]{@{}c@{}}LoC per\\Request $\times$100\end{tabular} \\ \midrule
Rand-\apirl\ & 88048 & 1019 & 14423 & 81.3\% &\gradientRR{6.575331} \\ \midrule
\apirl-r & 83756 & 1533 & 6930 & 87.2\% & \gradientRR{7.914464} \\
\apirl-u & 91597 & 1436 & 2805 & 92.4\% & \gradientRR{8.069743} \\
\apirl & 90079 & \textbf{1591} & 2805 & \textbf{95.6\%} & \gradientRR{8.469669} \\
\apirl-m & 90535 & 1562 & 2421 & 87.2\% & \gradientRR{7.721958} \\
\apirl-p & \textbf{100086} & 728 & \textbf{0} & 85.7\% & \gradientRR{7.115172} \\ 
\apirl-arat & 90222 & 1037 & 4531 &  90.7\% & \gradientRR{7.925243} \\ \midrule
\apirl-cov-r & 93748 & 5256 & 5987 & 92.4\% & \gradientRR{7.366231} \\
\apirl-cov-u & 92150 & 5499 & 7961 & \textbf{97.6\%} & \gradientRR{7.735177} \\
\apirl-cov & 87633 & \textbf{11197} & 6138 & 96.4\%  &\gradientRR{7.670852} \\
\apirl-cov-m & 91936 & 4562 & 3498 & 92.4\% & \gradientRR{7.734189} \\
\apirl-cov-p & \textbf{104979} & 11 & \textbf{11} & 90.8\% & \gradientRR{7.237988} \\ \bottomrule
\end{tabular}}
\end{table}

\paragraph{Effect of Reward Function}
\new{
The variations of \apirl-cov\ use a higher number of requests than \apirl\ equivalents, which reduces the coverage per request.
As \apirl\ models try to find bugs quickly they terminate episodes early, which in turn increases efficiency of coverage per request.
Such results confirm  \apirl-cov\ models \emph{indirectly} finding bugs by increasing coverage. 
%
%
Compared to learnt policies,} Rand-\apirl\ makes 108\% more invalid requests (4XX), and has the  worst coverage and coverage per request of any model.

\apirl\ and \apirl-cov-u achieve the highest code coverage for each reward function type (request based, and LoC). 
\new{Additionally, \apirl\ finds the most error requests (5XX) out of \apirl\ models, indicating both the breadth and depth of the learnt model. 
The reward based ratio (\apirl-r and \apirl-cov-r) achieves low coverage and successful requests (2XX) in both reward ablations.
This is likely due to diminishing returns, \ie\ the delta of a single step has small impact compared to the denominator as training progresses.
Uniform rewards for \apirl-u, \apirl-cov-u show the inability to differentiate between successful and error requests, resulting in high numbers of invalid requests and lower errored requests. 
Similarly, \apirl-arat's reward results in the model being unable to distinguish between finding bugs, and the undesirable behaviour of invalid requests.
%
\apirl-cov finds the most bugs of the ablations (89\% more than the next \apirl-cov\ variant). However, it achieves lower coverage in comparison to \apirl-cov-u, as \apirl-cov can still receive positive reward when recovering the same code.

}

\camera{
These results highlight the intricacies of reward functions, showing how even well considered rewards may not yield the optimal outcome. Furthermore, coverage-based rewards achieve better coverage compared to request-based.
\apirl\ is the best model in class, with more coverage, error requests, and LoC per request. 
}


\paragraph{Effect of Transformer}
The ablation of transformer embeddings in \apirl-m leads to an 8.4\% reduction in coverage. 
%
%
Similarly, \apirl-cov\ achieves a higher coverage compared to \apirl-cov-m. 
\new{
Thus, as in Table~\ref{tab:ablations}, transformer embeddings lead to the highest efficiency of LoC per request. 
The non-transformer variants perform comparably in terms of successful requests. However, both transformer based models find more bugs, with \apirl-cov finding the most errored requests of any model.
These results showcase the utility using the structured HTTP responses for feedback in learning. As policies learns to effectively maximise the learning objective (finding bugs, or increasing coverage).
%
}

\paragraph{Effect of RL training algorithm}
Using PPO leads to minimal invalid requests, and has the highest number of 2XXs. 
However, it finds the fewest bugs and the lowest coverage, 6.8\% and 9.9\% lower coverage than \apirl-cov\ and \apirl.
Upon manual inspection it is due to the PPO model entering local-optima, replicating action sequences, rarely deviating from these to maximises successful requests. 
\new{We speculate the off-policy nature of DQN learns a general mutation strategy, while PPO struggles to adjust over the curriculum. Specifically, the replay buffer in the DQN architecture provides a history of experiences, resulting in a greater degree of generalisation over the different \api\ endpoints.

}

\section{Related Work}\label{sec:related_work}


\new{

%

Diverse techniques have been used to test \api s.
Kim \etal~\shortcite{kim_enhancing_2023} enhance the OpenAPI specification via NLP techniques.
MINER~\shortcite{lyu_miner_2023} uses an attention based neural network to generate parameters.
RestTestGen~\shortcite{corradini_resttestgen_2022} traverses an ODG to develop call sequences.
EvoMaster~\shortcite{arcuri_automated_2021} presents an evolutionary algorithm that uses only the HTTP response in black-box settings to guide its fitness function.}
By comparison, \apirl\ parses direct feedback to a latent representation resulting in significantly more bugs in fewer requests.
\new{
\rat~\shortcite{kim_adaptive_2023} uses separate Q-tables to prioritise parameters and value-mapping functions when testing \api s. \rat\ also uses a reward based on response code, however our extensive experiments show that less granular rewards, such as those used by \rat\, can harm performance. Additionally, \apirl s deep architecture learns which parameters to include, how to manipulate values, and how to alter the HTTP method and auth token. 
}
\camera{
\deeper~\shortcite{corradini_deeprest_2024} uses a multi-learning architecture to test \api s. 
PPO based agents select different operations, and a multi-armed bandit \emph{per parameter} to select data values from a variety of different sources including: user provided data, previously seen data, and LLM generated data. 
Finally, they apply mutators to the selected data by random selection. 
\apirl's architecture differs significantly from this by using a single deep learning element to select parameters or apply mutations. 
\apirl\ also only requires training on a single \api\ and can then test other APIs, whilst \deeper\ requires training on each \api.}

Other RL approaches for fuzzing have used off-the-shelf architectures, to find bugs in software~\cite{bottinger_deep_2018,li_fuzzboost_2022,li_alphaprog_2022}
These works use similar rewards based around code coverage, expressing it as a function of \textit{how much} new coverage is achieved.
As our experiments suggest, this can lead to diminishing returns as the total coverage achieved increases, placing greater weight on the rewards achieved early in training. 
\apirl\ demonstrates the utility of consistent reward functions.

RL has also been used for automation tasks, often using simple heuristics~\cite{zheng_automatic_2021,li_alphaprog_2022,bottinger_deep_2018,li_fuzzboost_2022} or manually defined features~\cite{foley_haxss_2022,lee_link_2022}.
\new{RL has even been used to test GraphQL APIs for denial-of-service by McFadden \etal~\shortcite{mcfadden_wendigo_2024}.
}
%
Yet, unlike \apirl, these approaches are limited in the feedback they can use. 
Specifically, they are unable to a) use input of unbound length, b) use diverse input that contains subtleties relating to the test case, and c) generalise to unique, unseen request-templates without the need for retraining.

\camera{
\section{Limitations}

Two \apirl\ components require training: the RL agent, and the transformer. Any training set used will bias the final performance of the neural network. 
To limit bias and overfitting we use data from a diverse range of real \api s in the training set of \apirl s transformer.

We use a pre-trained element in the \apirl\ framework as transformers in RL can lead to training instability, requiring longer training, if they converge at all~\cite{parisotto_stabilizing_nodate}.
%
We also select an open-source example for training the policy element of \apirl. 
This is with the intention that a trained agent can be used to test other \api s without the high number of training iterations that are required to reach optimal performance. 
While larger, more complex \api s could be used, the empirical results demonstrate the performance of the agents from training on the didactic training set used.
We also do not test on live \api s due to ethical concerns over manipulation of API data or potential service outages. However, we use production grade \api s locally in an effort to bridge the gap.

The action set of \apirl\ constrained, in particular in regard to the ability to add key-value pairs to the request.
Other approaches can add targeted or guessable key-value pairs using known heuristics or generative methods~\cite{lyu_miner_2023}. 
Adding additional ways to generate key-value pairs for \apirl\ would be possible, at the cost of additional engineering effort.
As a result, given the performance of \apirl\ in Section~\ref{sec:experiments} we believe adding in this functionality serves limited purpose.
}

\section{Conclusions and Future Work}\label{sec:conclusion}

Testing \api s is key to ensuring the continued functioning of web infrastructure.
Thus, we propose a deep RL framework to test for bugs using a combined state representation from manual features and a pre-trained transformer.
Our implementation, \apirl, leverages complex and varied responses from \api s as feedback for learning.
We \new{conduct an extensive ablation study of rewards and design choices showing how they affect behaviour.}
%
We show that \apirl\ consistently achieves higher code coverage and finds more bugs than the SOTA, in a lower request budget.
Bugs we found have been reported and are either already fixed, or in the process of being fixed.

\camera{In future work, other approaches could add targeted or guessable key-value pairs using known heuristics or generative methods~\cite{lyu_miner_2023}. 
%
However, given \apirl s performance in Section~\ref{sec:experiments} we believe that functionality may serve a limited purpose.
%
Furthermore, RL approaches could be trained and tailored to specific APIs \eg\ GraphQL. 
We believe this to be interesting, and we have already shown the potential for generalisation of \apirl\ on 26 different \api s. 



}

%
%
%
%

\section*{Acknowledgements}
Research partially funded by EPSRC grant EP/T51780X/1 and the Defence Science and Technology Laboratory (Dstl). Dstl is an executive agency of the UK Ministry of Defence providing world class expertise and delivering cutting-edge science and technology for the benefit of the nation and allies. The research supports the Autonomous Resilient Cyber Defence (ARCD) project within the Dstl Cyber Defence Enhancement programme.

\clearpage
\newpage

\bibliography{references}

\clearpage
\newpage

\appendix

\section{Related Key-value pairs}\label{app:key-value-pairs}


We find related parameters to form good \api\ requests. 
%
%
Initially we iterate over all operations twice to capture all related key-value pairs. This is under the intuition that we may need to utilise some of these key-value pairs in order to gather additional parameters. If there are no dependencies we default to using randomly generated values that correspond to the parameter type, (\eg\ a string of 6 random characters, or random integers).

We then determine related key-value pairs by the common values seen in responses.
As an example, two separate endpoints may respond with parameters \code{username: alice1}, \code{name: Alice}, and \code{username: alice1}, \code{user\_id: 1}. We can then see that username \code{alice1} appears in both responses, thus we consider, the three key-value pairs to be related to each other.

While it would be possible to select the most preferable keys in certain settings (\eg\ \code{username} over \code{name}) we treat keys agnostically, as it is not possible to determine key preference in all settings, (\eg\ if keys such as \code{username}, \code{id}, \code{user\_id} were used in a \api\ response). 
Using the concrete values most recently seen, we improve the chances of using \emph{valid} related values. This allows us to be agnostic of the specific operations, this is important as there are many different combinations of parameters and operations that could occur.

\section{Hyperparameters}\label{app:hyperparam}

 Lower and upper bounds of the hyperparameters used in the grid search for the DQN and RoBERTa are displayed in Table~\ref{table:hyperparam}. Values were sampled uniformly to determine the optimal hyperparameters. We also include the selected values. 
 
\begin{table}[!ht]
        \centering
        \caption{Grid search ranges for hyperparameters neural network models\label{table:hyperparam}}
        \centering 
        {\fontsize{9}{10}\selectfont
\begin{tabular}{@{}llccc@{}}
\toprule
Model & Hyperparameter & \multicolumn{1}{l}{\begin{tabular}[c]{@{}l@{}}Lower \\ Bound\end{tabular}} & \begin{tabular}[c]{@{}c@{}}Upper \\ Bound\end{tabular} & Selected \\ \midrule
\multirow{5}{*}{DQN} & $\gamma$ & \multicolumn{1}{c}{0.5} & 0.999 & 0.9 \\
 & $\alpha$& \multicolumn{1}{c}{0.05} & 0.0005 & 0.005 \\
 & Batch Size & \multicolumn{1}{c}{32} & 256 & 128 \\
 & Update Step & \multicolumn{1}{c}{10} & 200 & 100 \\
 & Episode Length & \multicolumn{1}{c}{5} & 20 & 10 \\ \midrule
 \multirow{5}{*}{PPO} & $\gamma$ & \multicolumn{1}{c}{0.5} & 0.999 & 0.9 \\
 & $\alpha$& \multicolumn{1}{c}{0.05} & 0.0005 & 0.005 \\
 & Entropy Coef & \multicolumn{1}{c}{0.01} & 1.0 &  0.15 \\
 & Clip Coef &  \multicolumn{1}{c}{0.01} & 1.0 &  0.1 \\
 & Episode Length & \multicolumn{1}{c}{5} & 20 & 10 \\ \midrule
\multirow{2}{*}{RoBERTa} 
 & Batch Size & \multicolumn{1}{c}{32} & 256 & 64 \\
 & Training Epochs & 5 & 20 & 10 \\ \bottomrule
\end{tabular}}
        
\end{table}

\section{\api\ dataset}\label{app:data}

The \api\ dataset consists of 2566 
responses from 1283 operations, responses were collected from the \api s on April 29th 2023.
\api s were identified from an opensource OpenAPI Specficiation platform\footnote{\url{https://app.swaggerhub.com/search}}.
This dataset can be found in the additional source code appendix in the \code{APRIL/pre\_processing/api\_dataset.txt} file.
\api s used in the dataset range from popular cryptocurrency platforms such as coindesk, online dictionary services such as Wiktionary, image generation tools, and open-souce database services.
We save the HTTP response data from each endpoint collating it with its corresponding HTTP request.

This data was captured through the several steps:
\begin{enumerate}
    \item Crawling an opensource OpenAPI specification platform (\url{https://app.swaggerhub.com/search}) to extract OpenAPI specifications. This allows us to make use of publicly available \api s ensuring we do not collect any personal or private information.

    \item Manually verify that each of the \api s has a public facing endpoint that can be accessed. This further ensures that we are able to capture data from only live \api s.
    \item Using the OpenAPI specification and the \apirl\ framework for sending requests and collecting responses we send HTTP requests to each operation, capturing the responses. We also retain the parameter values from these responses in order to use their dependencies in making well formed requests.
    \item We then send a request to each of the endpoints a second time, using the previously seen parameter values. This improves how well-formed the requests are in more complex \api s \cite{martin-lopez_specification_2021}. Thus also improving the number of responses that contain meaningful information from which the RoBERTa model can learn.

\end{enumerate}

When training the RoBERTa model used in \apirl\ we preprocess each of the responses to form string based \emph{sentences} for input to the feature extractor, including: the status code, headers (removing those that contributed to noise in learning, such as \code{Date}, \code{etag}, \code{report-to}, or \code{Last-Modified}), encoding, HTTP protocol, and any redirects. These sentences are then used to create a vocabulary via Byte Pair Encoding to create a vocabulary of 52,000 tokens. The transformer is trained via using masked language modeling to learn the representation of REST API responses

\section{Example bug finding sequence}\label{app:example_bug}
\apirl\ performs a chain of mutations to find a bug in Spree Commerce's \code{api/v2/storefront/account} endpoint when trying to update a user through a series of \code{PATCH} requests. \apirl\ modifies the\code{user} parameter: removing the \code{user} object (Line 2), casting it as an empty string (Line 3), then adding a wild card (Line 4) that bypasses input sanitisation to cause an error when trying to access the attributes of the user.

\begin{listing}[tb]%
\caption{Mutations made by \apirl\ to parameter \code{user} that finds a bug in Spree Commerce.}%
\label{lst:listing}%
\begin{lstlisting}[language=json]
"user": {"email": "spree@example.com", "first_name": "John", "last_name": "Snow", "password": "spree123", "password_confirmation": "spree123"}
"user": {}
"user": ""
"user": "*"
\end{lstlisting}
\end{listing}

\begin{figure}
    \centering
    \includegraphics[width=0.9\linewidth]{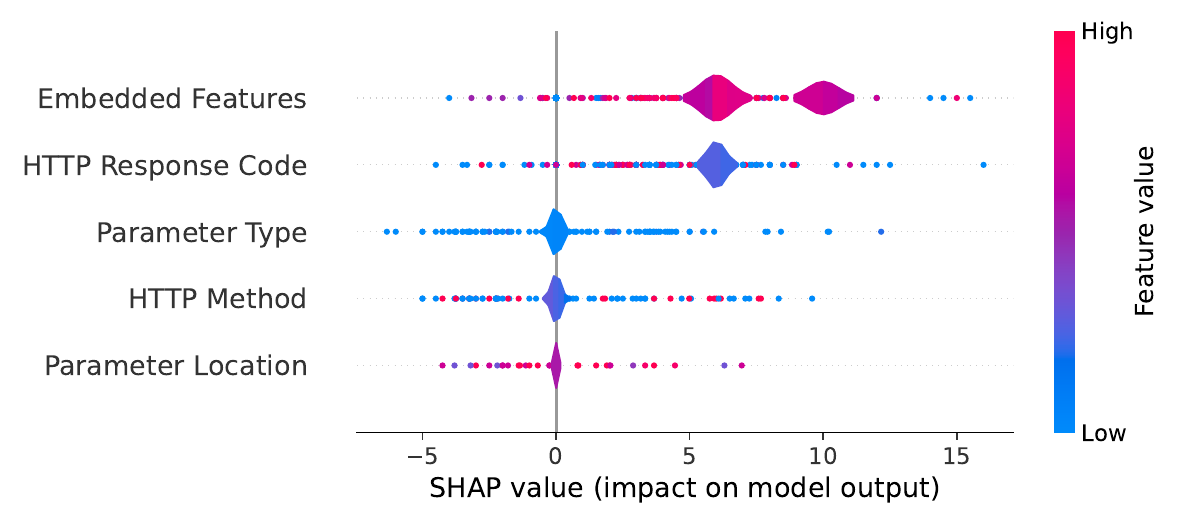}
\caption{SHAPley Values for all features in descending order of importance for \apirl.}     \label{fig:shapley_values}
\end{figure}

\section{Feature importance analysis}\label{app:importance}

We also investigate how \apirl\ makes use of the state representation 
via feature importance analysis.
%
We use the well known SHapley Additive exPlanations~\cite{lundberg_unified_2017} (SHAP) framework used in our feature importance analysis. It uses a game theoretic approach to determine the importance of input features on model outputs, doing so in an implementation agnostic way. SHAP connects the optimal credit allocations and local explanations to determine \textit{SHAPley values}, providing a way to determine the contribution of individual features within the complete feature space~\cite{lundberg_unified_2017}.

Specifically, we use the states seen in testing to compute SHAPley values representing the impact of each feature, and how their variance effects the RL agent. 
Figure~\ref{fig:shapley_values} shows how each variable influences \apirl, from most to least influential in descending order. 
%
\apirl\ shows a positive correlation (particularly in the highest ranking features), indicating the policy uses these optimally to create successful requests.
The deep embeddings from the transformer rank as most influential, indicating that this representation of additional response information, \eg\ `value missing' or `Example is invalid', provides meaningful granular feedback beyond a simple 400 HTTP code.
%
The HTTP response code is second with a wide distribution. 
Showing that the policy has learnt the response code is important to dictate strategy, correlating with 
previous works \cite{atlidakis_restler_2019,godefroid_intelligent_2020}.
The remaining features (Parameter Type, HTTP Method, and Parameter Location) then have a reducing distribution and influence.
This is due to the actions 
relating to certain actions, making them more situational than Embedded Features and HTTP Response code. 
\section{Coverage Collection}\label{app:code}
Coverage collection can be a non-trivial task depending on the system under test. We make use of standard libraries where possible to ensure consistently with prior work. In particular we use Xdebug \shortcite{noauthor_xdebug_nodate} for PHP and Jacoco \shortcite{noauthor_jacoco_2023} for Java.
For code collection in Python we use a hook function implemented by Atlidakis~\etal~\shortcite{atlidakis_restler_2019}. This is implemented using standard libraries \code{sys} and \code{functools}. While this is a custom hook, it was designed specifically for collecting coverage of Python based \api\ implementations.

\end{document}